\documentclass[aps, prd, a4paper, floatfix,showpacs, preprintnumbers, superscriptaddress, nofootinbib, onecolumn]{revtex4}
\usepackage{amssymb,amsmath,amsfonts,amsbsy,graphicx,rotating,bbold}
\usepackage{color,microtype}
\usepackage{tabularx}
\usepackage{eucal}
\usepackage{epsfig}
\usepackage{float}

\usepackage{xcolor}
\usepackage{lipsum}
\usepackage{textcomp}
\usepackage{siunitx}
\usepackage{makecell}
\usepackage{lipsum}

\usepackage{hyperref}
\usepackage[labelsep=period]{caption}
\DeclareCaptionJustification{justified}{\leftskip=0pt \rightskip=0pt \parfillskip=0pt plus 1fil}
\hypersetup{
	colorlinks=true,
	linkcolor=blue,
	filecolor=red,   
	citecolor= red,
	urlcolor=magenta}
\makeatletter
\let\ams@underbrace=\underbrace
\def\underbrace#1_#2{``
	\setbox0=\hbox{$\displaystyle#1$}``
	\ams@underbrace{#1}_{\parbox[t]{\the\wd0}{#2}}``
}
\makeatother

\graphicspath{{Images/}}

\usepackage{subcaption}
\definecolor{vividviolet}{rgb}{0.62, 0.0, 1.0}
\definecolor{amaranth}{rgb}{0.9, 0.17, 0.31}
\definecolor{palatinateblue}{rgb}{0.15, 0.23, 0.89}
\definecolor{brightpink}{rgb}{1.0, 0.0, 0.5}
\definecolor{cornflowerblue}{rgb}{0.39, 0.58, 0.93}
\definecolor{deepcarminepink}{rgb}{0.94, 0.19, 0.22}
\definecolor{radicalred}{rgb}{1.0, 0.21, 0.37}
\def\beq{\begin{equation}}
\def\eeq{\end{equation}}

\begin{document}

 	\title{Constraints on interacting dark energy models through cosmic chronometers and Gaussian process}
	
	\author{Muhsin Aljaf}
	\email{mohsen@mail.ustc.edu.cn}
	\affiliation{Department of Astronomy, University of Science and Technology of China, Hefei, Anhui 230026, China}
	\affiliation{Department of Physics, College of Education, University of Garmian, Kalar, Kurdistan region, Iraq}
	
	\author{Daniele Gregoris}
	\email{danielegregoris@libero.it}
	\affiliation{School of Science, Jiangsu University of Science and Technology, Zhenjiang 212003, China}
	\affiliation{Center for Gravitation and Cosmology, College of Physical Science and Technology, Yangzhou University,180 Siwangting Road, Yangzhou City, Jiangsu Province 225002, China}
	\affiliation{School of Aeronautics and Astronautics, Shanghai Jiao Tong University, Shanghai 200240, China}
	
	\author{Martiros Khurshudyan}
	\email{khurshudyan@ice.csic.es}
	\affiliation{Institute of Physics, University of Silesia, Katowice, Poland}
	\affiliation{Consejo Superior de Investigaciones Cientcas, ICE/CSIC-IEEC,
	Campus UAB, Carrer de Can Magrans s/n, 08193 Bellaterra (Barcelona) Spain}
\affiliation{International Laboratory for Theoretical Cosmology,Tomsk State University of Control Systems and Radioelectronics (TUSUR), 634050 Tomsk, Russia}

\begin{abstract}
	In this paper, after reconstructing the redshift evolution of the Hubble function by adopting Gaussian process techniques, we estimate the best-fit parameters for some flat Friedmann cosmological models based on a Modified Chaplygin Gas interacting with dark matter. In fact, the expansion history of the Universe will be investigated because passively evolving  galaxies constitute cosmic chronometers. An estimate for the present-day values of the deceleration parameter, adiabatic speed of sound within the dark energy fluid, effective dark energy, and dark matter equation of state parameters is provided. By this, we mean that the interaction term between the two dark fluids, which breaks the Bianchi symmetries, will be interpreted as an effective contribution to the dark matter pressure similarly to the framework of the \lq\lq Generalized Dark Matter". We investigate whether the estimates of the Hubble constant and of the present-day abundance of dark matter are sensitive to the dark matter - dark energy coupling. We will also show that the cosmic chronometers data favor a  cold dark matter, and that our findings are in agreement with the Le Ch\^atelier-Braun principle according to which dark energy should decay into dark matter.
\end{abstract}

\maketitle
\section{Introduction}
Despite being introduced for addressing a galactic puzzle, i.e., the flattening of the rotation curves, on cosmological scales dark matter combined together with dark energy can account for almost the full energy budget of the Universe. While there are still no experimental devices for confirming the existence of dark energy directly, the situation seems to be different for dark matter thanks to the model-independent study of its distribution within the Milky Way  \cite{iocco}. In the simplest cosmological scenario, the $\Lambda$ColdDarkMatter ($\Lambda$CDM) model,   dark matter is macroscopically pictured as a pressureless fluid \cite{peebles}.  However, different microscopic foundations for dark matter have been proposed linking it to some fundamental elementary particle theories like those of massive neutrinos \cite{massive}, sterile neutrinos \cite{sterile}, axions \cite{axions}, axinos  \cite{axino}, gravitinos  \cite{gravitino}, and neutralinos  \cite{neutralino}, just to mention a few examples (for a review of the different proposals of dark matter modelings in terms of elementary particles beyond the standard model, and how they affect the possible detection methods see  \cite{book1,book2}). However, massive neutrinos may not explain the formation of large-scale structures \cite{75,347}, while, on the other hand, sterile neutrinos and axions are consistent with the CP violation \cite{130,283}. Furthermore, a detection of dark matter constituted of axinos, gravitinos, or neutralinos can lead to an experimental confirmation of supersymmetric field theories \cite{susy}. Microscopically, the possible different modelings of dark matter can be classified into hot (with the massive neutrinos being one example), warm (as for sterile neutrinos), and cold  (like for axions and neutralinos) depending on the energy scale of the elementary particles constituting this fluid \cite{peebles}.

The aforementioned $\Lambda$ColdDarkMatter model assumes a dark energy fluid equivalent to a cosmological constant term entering the Einstein field equations, and that the two dark fluids are separately conserved, i.e., that they do not interact with each other through any energy exchange.  A consistent joint  interpretation of Planck results and weak lensing data however  suggests that some redshift evolution of the dark energy equation of state parameter may be necessary \cite[Sect.6.3]{planck}.  
Furthermore, interactions between dark energy and dark matter can alleviate the coincidence problem \cite{coincidence,coincidence1,coincidence2,coincidence3},  and  mitigate the discrepancies between the estimates of the Hubble constant from cosmic microwave background measurements or large scale structures versus supernovae data  as argued in \cite{referee1,referee2}; we refer as well to \cite{referee3,referee4,tension,tension1,tension2,tension3,tension4,tension5,tension6,tension7,tension8,tension9,tension10,tension11,tension12,tension13,tension14,Rodrigo3} for quantitative analyses on whether energy flows between dark matter and dark energy affect the estimates of various cosmological parameters.   Complementary studies have investigated how the growth of instabilities in interacting dark models affects the formation of astrophysical structures \cite{growth0,growth1,growth2,growth3,growth3a,growth3b,growth3c,growth4}, such as primordial black holes \cite{growth5,growth6,growth7}, and galactic halos \cite{halo1,halo2,halo3,halo4,halo5,halo6,halo7,halo8,halo9,halo10,halo11} due to the fact that the density of dark matter does not dilute anylonger with the cube of the scale factor of the universe. Gravitational waves has been used for constraining dark interactions as well \cite{siren1,siren2,siren3,siren4}.  
Moreover, a coupling  between the dark energy field and dark matter, with the latter pictured as neutrinos, affects the neutrinos' masses estimates \cite{neutrino1,neutrino2, neutrino3}. From a more mathematical point of view, specific interplays between the equation of state of dark energy and the interaction term with dark matter can give rise to different types of finite-time kinematic and matter density singularities \cite{ss6,symmetry2021}.

For taking into account the observational requirement of an evolving equation of state of dark energy, we model the dark energy fluid as a Modified Chaplygin Gas \cite{MCG,Sergei}, rather than considering just a redshift parametrization \cite{par,par1}, because of its well established physical motivation. In fact, this fluid approach belongs to the wider class of chameleon field theories in which the constant equation of state parameter $p=w\rho$ is promoted to an energy-dependent functional according to $w \to w(\rho)$, and therefore it exhibits a sort of running \cite{cham1,cham2}. 
In particular, our fluid model interpolates between an ideal fluid behavior at low energy densities and a Generalized Chaplygin Gas in the high energy limit. Therefore, we implement a sort of {\it asymptotic freedom} at low energies because the interactions within the fluid are suppressed \cite{gross1,gross2}, while at high energies, we match with the Born-Infeld paradigm with our model being formulated in terms of the Nambu-Goto string theory  \cite{bento}.

Therefore, in this paper, we test a set of dark energy - dark matter interacting models with the purpose of enlightening the physical properties of dark matter. In fact, the interaction term between the two fluids behaves as an {\it effective pressure} entering the energy conservation equation, and consequently affecting the dust picture of dark matter. Thus, evaluating the effective equation of state parameter for the dark matter, we can discriminate between cold, warm, and hot models.

From the technical point of view, we employ Gaussian Process techniques for reconstructing the redshift evolution of the Hubble function with the purpose of selecting the best cosmological model involving energy flows between dark matter and dark energy. The latter is modeled in the form of the Modified Chaplygin gas. In particular, we use 30 data points for $H= H(z) $  consisting of samples deduced from the differential age method, allowing the Gaussian Process to constrain the model parameters. Our purpose is to extend and complement the analysis of \cite{Melia,sar} by allowing a redshift-dependent equation of state for dark energy (for accounting for Planck observations), and interactions in the dark sector (for alleviating the coincidence problem).

Our paper is organized as follows: we introduce our cosmological model in Sect. \ref{II} reviewing the physical properties of the Modified Chaplygin Gas and the features of the postulated energy exchanges between the two cosmic fluids. Then, in Sect. \ref{III}  we explain the importance of the cosmic chronometers as model-independent observational data for the reconstruction of the Hubble function, and for constraining the values of the free parameters entering our class of models. In Sect. \ref{IV} we present the reconstruction for the Hubble function through gaussian processes, while in Sect. \ref{IVb} we describe the numerical method we have adopted for the integration of the field equations. The same Sect.  exhibits explicitly also our cosmological results comparing and contrasting between the different possible choices of the interaction term. Lastly, we conclude in Sect. \ref{V}  with some remarks about the importance of our study in light of the current literature estimates of the cosmological parameters by means of various different datasets. 

\section{Overview of the cosmological model}\label{II}

In this section we will introduce the basic equations of the cosmological model under investigation. For the geometrical modeling of the Universe we adopt the flat Friedmann metric which, in a Cartesian system of coordinates, reads \cite{exact}: 
\begin{equation}
d s^{2}=-d t^{2}+a^{2}(t)(dx^2 + dy^2 +dz^2)\,,
\end{equation}
where $a(t)$ is the time-dependent scale factor of the Universe. Moreover, we picture the matter content of the Universe as two perfect fluids with energy density $\rho(t)$ and pressure $p(t)$, respectively. Their stress-energy tensors are $T^{\mu}{}_{\nu}={\rm diag}[-\rho_i(t), p_i(t), p_i(t), p_i(t)]$ with $i=de\,, m$ for dark energy and dark matter respectively.  The relevant Einstein field equation $G_{\mu\nu}=8 \pi G\, T_{\mu\nu}$  is given by
\begin{equation}\label{Fridm}
\left(\frac{\dot a }{a}  \right)^2:=H^{2}=\frac{1}{3 M_{p}^{2}}\left[\rho_{d e}+\rho_{m}\right]\,,
\end{equation}
where $M_{p}^{2}=(8 \pi G)^{-1}$ is the reduced Planck mass, $H$ is the Hubble function, and an overdot denotes a time derivative. Then, the Bianchi identities $T^{\mu\nu}{}_{;\nu}=0$ deliver 
\begin{equation}
\begin{array}{l}{\dot{\rho}_{m}+3 H \rho_{m}=0}\,, \\ {\dot{\rho}_{d e}+3 H\left(\rho_{d e}+p\right)=0\,,}\end{array}
\end{equation}
which account for two separately-conserved dark matter and dark energy fluids. However, in this paper we will introduce an interaction term $Q$ between these two fluids breaking the Bianchi symmetry (of course the total energy of the Universe is still conserved because dark matter is transformed into dark energy or viceversa), and the coupled evolution of the two fluids is now given by
\begin{equation}\label{cons}
\begin{array}{l}{\dot{\rho}_{m}+3 H \rho_{m}=Q}\,, \\ {\dot{\rho}_{d e}+3 H\left(\rho_{d e}+p\right)=-Q\,.}\end{array}
\end{equation}

\subsection{Modeling of dark energy as a Modified Chaplygin Gas}

For the modeling of the dark energy fluid we adopt the Modified Chaplygin Gas proposal based on the equation of state \cite{GCG}:
\begin{equation}\label{chapeq}
p=A \rho_{de}-\frac{B}{\rho_{de}^{\alpha}}\,,
\end{equation} 
in which $A$, $B$ and $\alpha$ are constant parameters while $\rho_{de}$  is the energy  density of the fluid. The modified version of the Chaplygin gas is an extension of the Generalized Chaplygin gas whose limit corresponds to the choices $ A = 0  $ and $ \alpha > 0 $; also, selecting $A=0$ and $\alpha = 1$ the model  reduces to the original Chaplygin gas.
The Modified Chaplygin Gas implements a form of {\it effective freedom} in the cosmic fluid \cite{gross1,gross2}. In fact, if $\alpha>0$ then the equation of state (\ref{chapeq})  reduces to that of an ideal fluid with pressure and energy density directly proportional to each other $p \propto \rho$ at high energies (which can possibly occur in the first instants after the big bang). On the other hand, if $\alpha<0$ the linear behavior is realized at low energies (i.e., at late ages) when the fluid dilutes due to the expansion of the Universe.  Since the constituents of an ideal gas have only kinetic and not potential energy, in these two regimes they essentially behave as free particles.  The occurrence of one of these two cases will be explored in this paper through the use of the cosmic chronometers. The Modified Chaplygin Gas has been tested in \cite{constraint1,constraint2,constraint3,constraint4} against Constitution$+\mathrm{CMB}+\mathrm{BAO}$ data, and against Union$+\mathrm{CMB}$ $+$ BAO observations using Markov Chain Monte Carlo techniques. In this paper, we will quantify the role of the interaction terms on the estimates of the cosmological parameters comparing with these literature results. More formally, exploiting the fluid - scalar field correspondence in the canonical framework \cite{corr1,corr2}, the pressure and energy density of the Chaplygin gas (\ref{chapeq}) can be related to the kinetic energy $X=-\frac12 g^{\mu\nu} \partial_\mu \phi \partial_\nu \phi$ and the potential $V$ of a scalar field $\phi$ via:
\beq
\rho_{de}=\frac{\dot \phi^2}{2} +V\,, \qquad p=\frac{\dot \phi^2}{2} -V\,,
\eeq
or equivalently
\beq
V=\frac12 \left[(1-A) \rho_{de} +\frac{B}{\rho_{de}^\alpha} \right]\,, \qquad \dot \phi^2= (1+A)\rho_{de} -\frac{B}{\rho_{de}^\alpha}\,.
\eeq
\cite{chappot} has extensively investigated the characteristics of the potential $V=V(\phi)$ in a flat Friedmann Universe dominated by the Modified Chaplygin Gas. 
At early times, which correspond to $a(t)\to0$ the potential can either approach zero (for $A=1$), or diverge (for $A\neq1$). At late times, which correspond to $a(t)\to\infty$, the potential approaches the constant value $V(\phi)\to \left(\frac{B}{1+A} \right)^{\frac{1}{1+\alpha}}$. Therefore,  at late times for $A=-1$ the potential diverges if  $\alpha>-1$, and approaches zero otherwise.
Analitically, in a flat Friedmann universe whose only energy-matter content is the Modified Chaplygin Gas (\ref{chapeq}) the potential of the underlying self-interacting scalar field is \cite{chappot,chappot1}:     
\begin{equation}
V(\phi)= \frac{1-A}{2}\left( \frac{B}{1+A} \right)^{\frac{1}{1+\alpha}}\cosh^{\frac{2}{1+\alpha}}\frac{\sqrt{3(1+A)}(1+\alpha)\phi}{2}+\frac{B}{2}\left( \frac{B}{1+A} \right)^{-\frac{\alpha}{1+\alpha}}\cosh^{-\frac{2}{1+\alpha}}\frac{\sqrt{3(1+A)}(1+\alpha)\phi}{2}\,.
\end{equation}

\subsection{Modeling of the interaction terms}

An interaction term between the dark matter particles  and the dark energy molecules behaves phenomenologically as an effective pressure $\Pi$  which couples  the conservation equations of the two cosmic fluids (breaking the Bianchi identities).  In general,  the interaction term would be written as \cite{rev1,rev2}
\beq
Q=3 H \Pi \,.
\eeq
In this paper, we consider an effective pressure parametrized as
\beq
\label{parpi}
\Pi=b \frac{\rho^m_{de} r^{-n}+(-1)^s \rho^m_{m} r^n}{\rho^u}\,,
\eeq
where the parameter $b$ quantifies the strength of interactions between dark energy and dark matter. This quantity cannot be fixed by any theoretical first principle argument,  and therefore its value will be estimated through the model selection procedure. A non-zero value for $b$ can be interpreted as a manifestation of a fifth force mediated by a postulated {\it cosmon field} acting between dark matter and dark energy  \cite{cosmon}  violating the weak equivalence principle \cite{fifth}. In particular, a positive $b$ implies that dark energy is decaying into dark matter, while a negative sign is consistent with an energy flow in the opposite direction. The Le Ch\^atelier-Braun principle favors a decay of dark energy into dark matter (and not viceversa) for maintaining the whole system close to thermal equilibrium because in this case the entropy of the universe will increase \cite{thermo1}. Interestingly, it seems that it is still an open question how to reconcile thermodynamically viable interacting models with the problem of formation of astrophysical structures \cite{salvatelliprl}. In (\ref{parpi})  $\rho=\rho_{de}+\rho_{m}$ is the total energy budget of the universe, and 
\beq
r=\frac{\rho_{de}}{\rho_{m}}
\eeq
is the relative abundance of the two cosmic fluids. In this section,  we will show that the elegant parametrization (\ref{parpi}) is rich enough for covering both the models with linear and nonlinear energy interactions,  and with fixed or variable direction of the energy flow. In fact, for $s$ odd the effective pressure $\Pi$ is allowed, at least in principle, to switch its sign during the time evolution of the Universe depending on the interplay between the densities of the two dark fluids. This scenario would correspond to a phase transition between decelerating-accelerating (or viceversa) phases of the universe \cite{sign1,sign2,sign3}. In light of this dependence on the background energy density  when $s$ is odd, the effective pressure can be interpreted as a {\it chameleon field}  \cite{cham1,cham2}.  We stress  that the interaction term (\ref{parpi})  relies only on the abundance of the two dark fluids, and not on their physical nature or modeling (we just need to assume a time-evolving dark energy, which therefore rules out the case of a cosmological constant) \cite{hoff}. To summarize, we can speak of {\it effective pressure} because in this class of models dark matter is behaving  as a non-ideal fluid with equation of state parameter $\Pi/\rho_{m}$, and not anylonger as pressureless dust \cite{weff}.  The inferred value for the effective dark matter equation of state parameter will allow us to discriminate between the models of cold vs. warm vs. hot dark matter.  The explicit models we will test in this paper are: 
\begin{equation}
\label{Q1}
Q_{1}=3 H b \rho_{de}\,,
\end{equation}
which corresponds to the choice $m=1$, $n=0$, $s \to \infty$, $u=0$.
\begin{equation}
\label{Q2}
Q_{2}=3 H b \rho_{m}\,,
\end{equation}
which corresponds to the choice $m=1$, $n=1$, $s \to \infty$, $u=0$.
\begin{equation}
\label{Q3}
Q_{3}=3 H b\left(\rho_{de}+\rho_{m}\right)\,,
\end{equation}
which corresponds to the choice $m=1$, $n=0$, $s=0$, $u=0$. 
\begin{equation}
\label{Q4}
Q_{4}=3 H b\left(\rho_{de}-\rho_{m}\right)\,,
\end{equation}
which corresponds to the choice $m=1$, $n=0$, $s=1$, $u=0$. In this model the effective pressure $\Pi$ may switch its sign during the evolution of the universe depending on the relative abundance between the two dark fluids, inverting the direction of the energy flow from dark matter to dark energy.
\begin{equation}
\label{Q5}
Q_{5}=3 H b \sqrt{\rho_{de} \rho_{m}}\,,
\end{equation}
which corresponds to the choice $m=1/2$, $n=1/2$, $s \to \infty$, $u=0$.
\begin{equation}
\label{Q6}
Q_{6}=3 H b \frac{\rho_{de} \rho_{m}}{\rho_{de}+\rho_{m}}\,,
\end{equation}
which corresponds to the choice $m=2$, $n=1$, $s \to \infty$, $u=1$. 
\begin{equation}
\label{Q7}
Q_{7}=3 H b \frac{\rho_{de}^{2}}{\rho_{de}+\rho_{m}}\,,
\end{equation}
which corresponds to the choice $m=2$, $n=0$, $s \to \infty$, $u=1$.
\begin{equation}
\label{Q8}
Q_{8}=3 H b \frac{\rho_{m}^{2}}{\rho_{de}+\rho_{m}}\,,
\end{equation}
which corresponds to the choice $m=2$, $n=2$, $s \to \infty$, $u=1$.

The interaction terms $ Q_{3} $ and $ Q_{4} $  are symmetric under the reflection $\rho_{de}\leftrightarrow \rho_{m}$. The linear interaction terms $ Q_{1} $-$ Q_{4} $  can be interpreted as a first-order Taylor expansion, which holds at low energy densities for any parameterization of the term $Q$. On the other hand, an analogy with chemical and nuclear reactions suggests that the interaction term should depend on the product of the abundances of the two species \cite{chemical}.  Lastly, looking at the Friedmann equation (\ref{Fridm}), we note that, as for any model with interactions, the evolution of the Hubble function remains decoupled from the evolution of the cosmic fluids \cite{decoupling}. We stress that when assuming these types of interactions, the following hypothesis should be made: interactions are negligible at high redshifts growing in strength at lower redshifts,  motivating the analysis of their impact on the cosmological parameters from available observational datasets.

\section{Cosmic chronometers data}\label{III}

The role of passively evolving early galaxies as {\it cosmic chronometers} permits to measure the expansion history of the Universe directly without the need of relying on any cosmological model, and in particular without the need of making any at {\it a priori} hypothesis on the nature of dark energy and dark matter. In fact,  this approach is based on the measurement of the differential age evolution as a function of the redshift for these galaxies, which in turn provides a direct estimate of the Hubble parameter:
\begin{equation} \label{DvsH}
H(z)=-\frac{1}{(1+z)} \frac{d z}{d t} \approx-\frac{1}{(1+z)} \frac{\Delta z}{\Delta t}\,.
\end{equation}
The redshift is related to the scale factor of the Universe via 
\begin{equation}
1+z\,=\,\frac{1}{a}.
\end{equation}
The dependence  on the measurement of a differential quantity, that is $\Delta z / \Delta t$, is the most important strength of this approach because it provides many advantages in minimizing some common sources of uncertainty and systematic effects (for a detailed discussion see \cite{Holsclaw}). We exploit $30$ data points of $H=H(z)$ consisting of 30 point samples deduced from the differential age method. Keeping this in mind, first, we will use Gaussian Process techniques for reconstructing the Hubble vs. Redshift evolution, and then we will optimize the free parameters of our family of cosmological interacting models. The data points we will consider are taken from \cite{data} and are exhibited in Table \ref{tabledata}. Then, we can select the best model by estimating the differential area $ \Delta A^{\prime} $ as explained in detail in the next section.

\begin{table}
	\begin{tabular}{|c|c|c|c|c|c|}
		\hline
		$ z $       & $ H(z) $ & $\sigma_{H}$  & $ z $      & $ H ( z ) $ & $\sigma_{H}$ \\ \hline
		0.07    & 69      & 19.6  & 0.4783 & 80.9    & 9     \\ \hline
		0.09    & 69      & 12    & 0.48   & 97      & 62    \\ \hline
		0.12    & 68.6    & 26.2  & 0.593  & 104     & 13    \\ \hline
		0.17    & 83      & 8     & 0.68   & 92      & 8     \\ \hline
		0.179   & 75      & 4     & 0.781  & 105     & 12    \\ \hline
		0.199   & 75      & 5     & 0.875  & 125     & 17    \\ \hline
		0.2     & 72.9    & 29.6  & 0.88   & 90      & 40    \\ \hline
		0.27    & 77      & 14    & 0.9    & 117     & 23    \\ \hline
		0.28    & 88.8    & 36.6  & 1.037  & 154     & 20    \\ \hline
		0.352   & 83      & 14    & 1.3    & 168     & 17    \\ \hline
		0.3802  & 83      & 13.5  & 1.363  & 160     & 33.6  \\ \hline
		0.4     & 95      & 17    & 1.4307 & 177     & 18    \\ \hline
		0.4004  & 77      & 10.2  & 1.53   & 140     & 14    \\ \hline
		0.4247  & 87.1    & 11.1  & 1.75   & 202     & 40    \\ \hline
		0.44497 & 92.8    & 12.9  & 1.965  & 186.5   & 50.4  \\ \hline
	\end{tabular}
	\caption{The observational data for $ H=H(z) $ and their uncertainty $\sigma_{H}$ in units of  km/s/Mpc. The 30 data points were obtained from the differential age method of cosmic chronometers. This table is taken from \cite{data} (see references therein for comments about each data point).}
	\label{tabledata}
\end{table}

\section{Gaussian process techniques for the $ H= H(z)$ Reconstruction}\label{IV}
Gaussian process techniques,  which have been studied in detail in \cite{Seikel}, constitute a set of model-independent algorithms that can be exploited for the reconstruction of the Hubble parameter; they are particularly useful when studying dark energy - dark matter interacting models.  This procedure relies on the following assumptions. First, it is assumed that each observational datum satisfies a Gaussian distribution in such a way that the full set of observational data obey to a multivariate normal distribution. The relationship between two different data points is accounted for by a function called {\it covariance function}. The values of the data at some redshift point at which they have not been directly measured would be extrapolated with the use of the covariance function because the points obey to the multivariate normal distribution. Besides, also the derivative (up to some order) of the function, that we want to reconstruct, at these data points,  can be calculated through the covariance function. Therefore, this mathematical formalism allows us to numerically reconstruct every smooth function at any point via its dependence on the data and the values of the slopes at those points. Thus, the crucial task in  Gaussian process techniques is to determine the covariance function at different points starting from the available measured data.

In general, when reconstructing a mathematical function through a gaussian process algorithm, different functional behavior of the covariance function may be implemented. The most convenient choice is to consider the probability distribution of the measured data points keeping in mind that the Gaussian process should be regarded as  a generalization of the Gaussian probability distribution.  In this paper, the observational data are the distances $D$ to the host galaxies which obey to a Gaussian distribution with certain known mean and variance. With this information in hand, Gaussian processes allow us to reconstruct at posteriorly the distribution of the function $ H(z)$  implementing the known Gaussian distribution characterizing $D$ into (\ref{DvsH}).

Therefore, the key of this algorithm is the covariance function $k(z_{1}, z_{2})$ which correlates the values of the distance to a certain galaxy $ D(z)  $ at the two different redshift ages $ z_{1} $ and $z_2  $. In general, one can choose from different functional behaviors for the covariance function $k(z_{1}, z_{2})$, all of which are characterized by the two hyperparameters  $\sigma_{f}$ and $\ell$; the latter would be determined testing against the observational data via a marginal likelihood. As a subsequent step, exploiting the inferred covariance function, the values of the function we want to reconstruct can be extrapolated at any arbitrary redshift point for which no measured data are available. Then, using the relation between the Hubble function $ H(z) $  and the distance $ D $, the redshift evolution of the Hubble function can be provided. Due to its model independence, this method has been widely applied in the reconstruction of dark energy equation of state and of the Hubble parameter \cite{Holsclaw1,Seikel1,Seikel2,Rodrigo1}, or in the test of the concordance model \cite{Busti,Cai,Rodrigo2},  for the analysis of the dynamical features of the dark energy by taming the matter degeneracy \cite{sign1}, and in light of cosmic chronometers in the $\Lambda$CDM model \cite{Melia}. The purpose of the present work is exactly to improve the latter by considering an evolving dark energy equation of state based on the Modified Chaplygin Gas fluid.

\begin{figure}[H]
	\centering
	\includegraphics[width=0.49\linewidth]{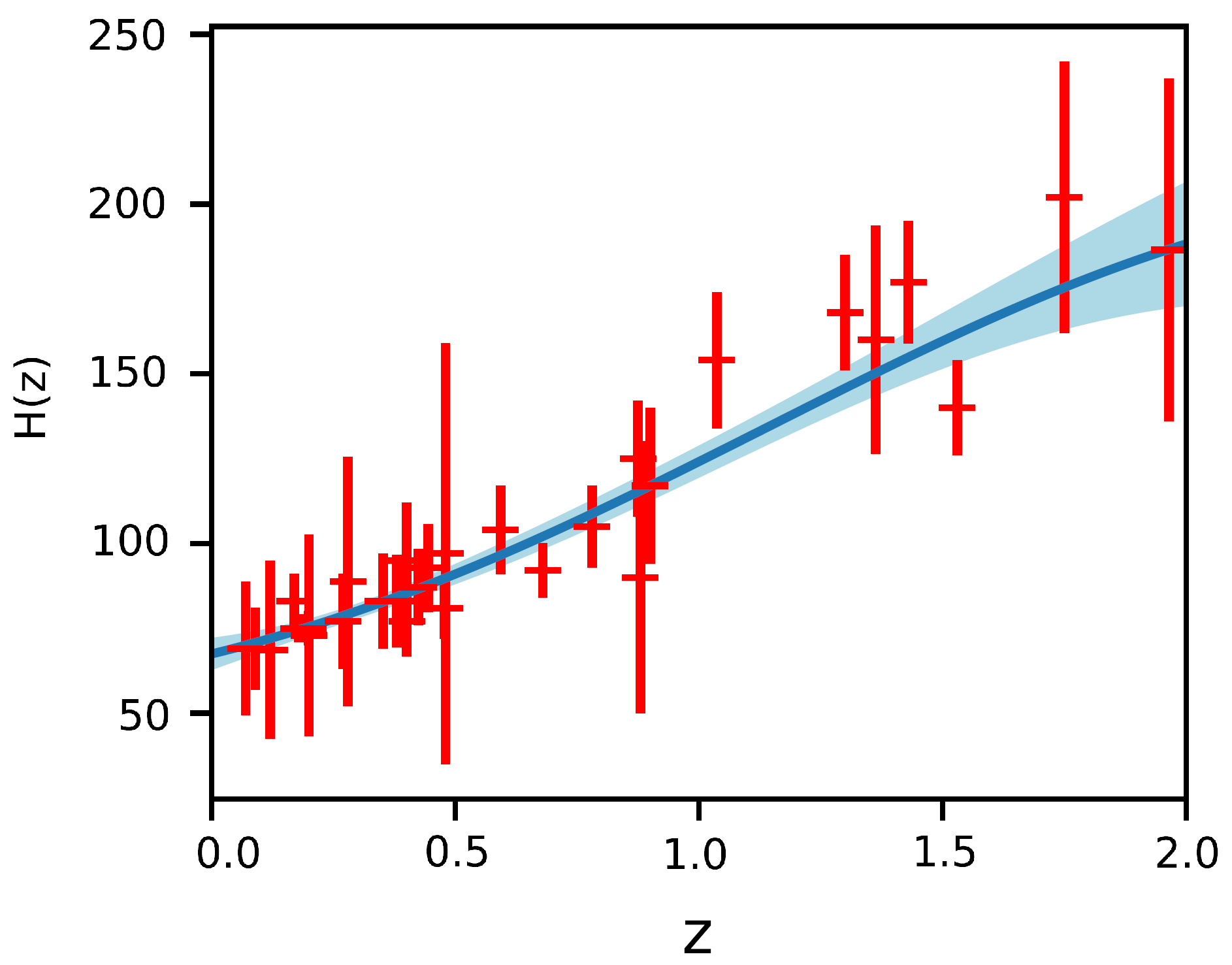}
	\caption{ The figure displays the reconstructed curve for $H=H(z)$ using Gaussian process techniques assuming an exponential covariance function and the 30 model-independent measurements of  $ H(z) $  with the corresponding error bars. The blue surface represents the 1$\sigma$ confidence region of the reconstruction.  }
	\label{H}
\end{figure}

In this paper we adopt a gaussian exponential distribution as our covariance function $k(z_{1}, z_{2})$:
\begin{equation}
\label{gkernel}
k\left(z_{1}, z_{2}\right)=\sigma_{f}^{2} \exp \left(-\frac{\left(z_{1}-z_{2}\right)^{2}}{2 \ell^{2}}\right)\,.
\end{equation}
We can reconstruct the redshift evolution of the Hubble function and that of the equation of state of dark  energy (Modified Chaplygin Gas in our case) by modifying the {\it GaPP} package  developed in \cite{Seikel}. We exhibit the outcome of the Reconstruction Process in Fig. \ref{H} in which we display both the reconstructed $H$ vs. $z$ curve and the 30 model-independent measurements of  $ H(z) $  with the corresponding error bars we have used (compare with Table \ref{tabledata}). The blue surface represents the 1$\sigma$ confidence region of the reconstruction.


\section{Numerical Analysis}\label{IVb}

We integrate the system constituted by the Friedmann equation (\ref{Fridm}) and by the energy conservation equations (\ref{cons}) using the iterative numerical differential equations solvers known  under the name of  {\it  Runge-Kutta method} \cite{RK}. This method uses the input for the initial values, let us  say ($ x_{n} $, $ y_{n} $),  for evolving them into   ($x_{n+1}$, $ y_{n+1} $) by use of a discretized system of equations. Explicitly, the steps of the numerical algorithm  we used to integrate our system of differential equations read as \cite{RK}:
\begin{equation}
\begin{aligned}
&K{1}=h \cdot f\left(x_{n}, y_{n}\right),\\
&K_{2}=h \cdot f\left(x_{n}+\frac{h}{2}, y_{n}+\frac{K_{1}}{2}\right),\\
&K_{3}=h \cdot f\left(x_{n}+\frac{h}{2}, y_{n}+\frac{K_{2}}{2}\right),\\
&K_{4}=h \cdot f\left(x_{n}+h, y_{n}+K_{3}\right),\\
&y_{n+1}=y_{n}+\frac{K_{1}}{6}+\frac{K_{2}}{3}+\frac{K_{3}}{3}+\frac{K_{4}}{6}\,,
\end{aligned}
\end{equation}
where $h$ is the step size and $ f(x,y) $ is the differential equation to solve, i.e.  the Friedmann equation and energy conservation equations, respectively.
Once provided with a set of initial conditions, this algorithm is able to deliver the values for the  Hubble function, the energy density of dark matter and of dark energy for each interacting model. After integrating numerically the theoretical field equations, we  implement the procedures from \cite{Seikel2} for the gaussian reconstruction, and from \cite{Melia} for the model selection:

\begin{itemize}
	\item We use the data from Table \ref{tabledata} to generate the mock samples for the 30 values of the Hubble function at the same redshift, and for each redshift value $z_{i}(i=1, \dots, 30)$, assuming that the measurements follow a  Gaussian randomized distribution:
	
	\begin{equation}
	H^{\operatorname{mock}}\left(z_{i}\right)=H\left(z_{i}\right)+r \sigma_{i}\,,
	\end{equation}
	
	where $ r $ is a Gaussian random variable with mean $0$,  variance $1,$ and $\sigma_{i}$ is the dispersion at $z_{i}$.
	
	\item  Then, we reconstruct the mock function $ H^{mock}(z) $, and we calculate a normalized absolute area difference between this function and the actual function using  the formula
	
	\begin{equation}\label{diffA}
	\Delta A^{\prime} =\frac{\int_{0}^{2} d z\left|H^{\operatorname{mock}}(z)-H(z)\right|\ }{\int_{0}^{2} d z H(z)} \,.
	\end{equation}
 The probability that the theoretical prediction of our cosmological model differs from the reconstructed function is quantified by the differential area $\Delta A^\prime$   which should be minimized by optimizing appropriately the values of the model free parameters. In fact, we will need to  estimate the possible randomized realizations which come with a differential area smaller than a specific value by presenting the cumulative probability distribution versus $\Delta A^\prime$. In our paper we are required to adopt this so-called {\it Area Minimization Statistic} rather than discrete sampling statistics, e.g. weighted least squares, because we are comparing two continuous curves and not isolated points	\cite[Sect.5]{area}.
	
	\item  Lastly, we build the distribution of frequency versus differential area $ \Delta A^{\prime} $  from
	which we can construct the cumulative probability distribution.
\end{itemize}

Applying this procedure to every interacting model  $ Q_{i} $ that we have introduced in Sect. \ref{II}, we calculate the differential area $ \Delta A^{\prime} $ from (\ref{diffA}) by replacing $ H_{mock} $ with the reconstructed function $ H_{i}(z) $.  Furthermore, we optimize the values of the free parameters characterizing each model (three parameters (A, B, $\alpha$) which enter the equation of state of the Modified  Chaplygin Gas as from (\ref{chapeq}), and the parameter  $b$ quantifying the strength of  the interactions between dark energy and dark matter). We allow these free paramaters to take values in the following ranges: $H_{0}$ $\in$ (40, 90), $\Omega_{m0}$ $\in$ (0.2, 0.7), A $\in$ (-2, 2), B $\in$ (-2, 2), b $\in$ (-1, 1), and  $\alpha$ $\in$(-1.0 , 1.0).

\subsection{Numerical results}

 Being the cosmic chronometers data in Table \ref{tabledata} dependent on the redshift and not on the time, for tackling the optimization process it is mathematically convenient to recast the model equations (\ref{Fridm})-(\ref{cons})-(\ref{chapeq}) as
\beq
\frac{dH}{dz}=\frac{3 H^2 +p}{2(1+z)H} \,,\quad
\frac{d \Omega_m}{dz}= -\frac{Q+3H \Omega_m p}{3(1+z)H^3} \,, \quad
p= 3 A(1-\Omega_m)H^2 -\frac{B}{(3(1-\Omega_m)H^2)^\alpha}\,,
\eeq
where we have used (\ref{DvsH}), the definitions of the matter parameters $\Omega_{de}=\frac{\rho_{de}}{3H^2}$ and $\Omega_m=\frac{\rho_m}{3H^2}$, and the Friedman equation $\Omega_{de} + \Omega_m=1$. Each interaction term $Q_i$ should also be re-expressed as a function of $\Omega_m$ and $H^2$, rather than of $\rho_m$ and $\rho_{de}$, accordingly. 

\begin{figure}[ht]
	\centering
	\includegraphics[width=0.6\linewidth]{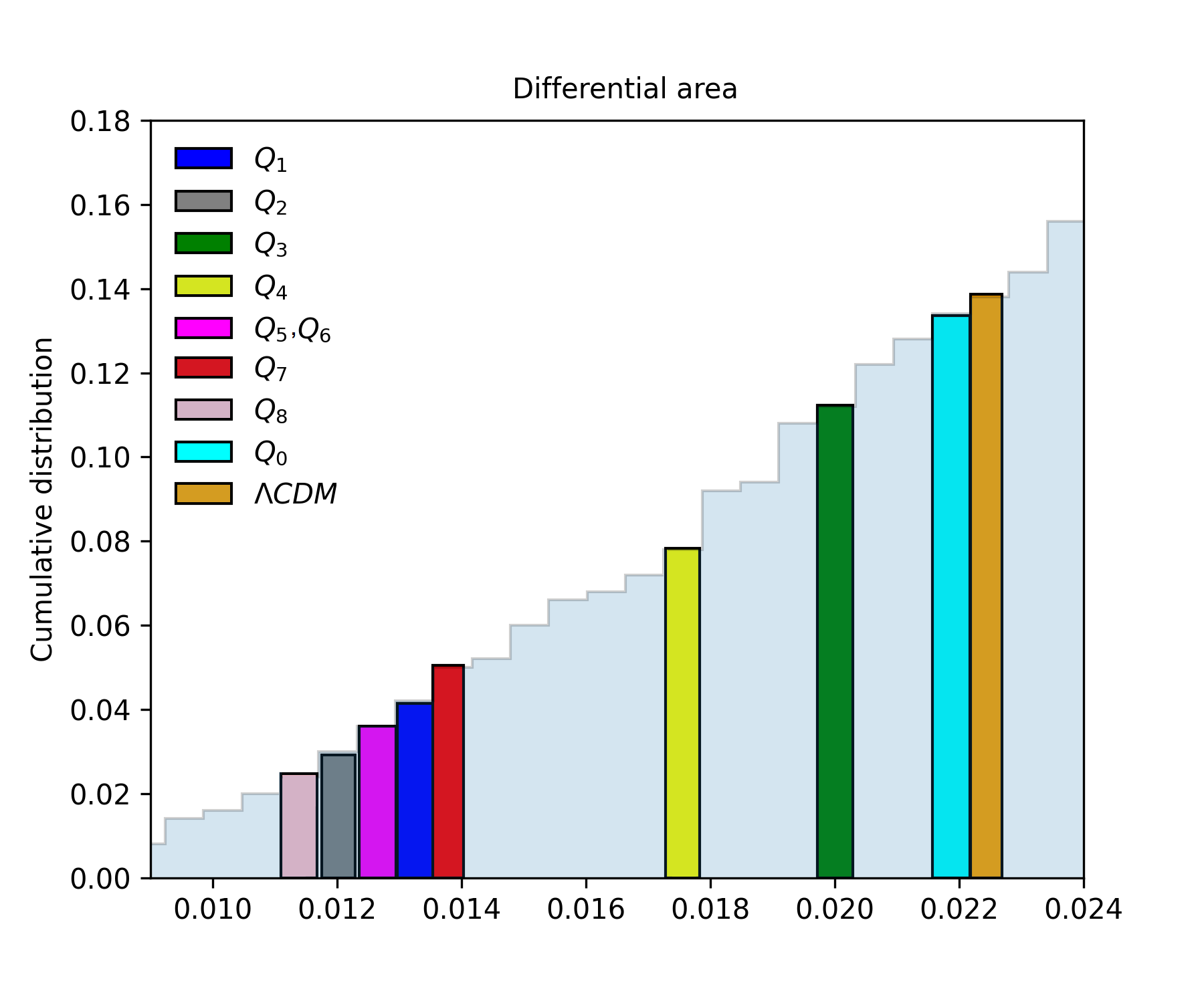}
	
	\caption{This figure depicts the cumulative distribution of the differential area $\Delta A^{\prime} $, as calculated from (\ref{diffA}), and whose numerical results are reported in Table \ref{table33} together with the optimal values for the model free parameters.  The specific forms of the dark energy - dark matter interaction terms  $Q_1$, ..., $Q_8$ can be found in eqs. (\ref{Q1}), ..., (\ref{Q8}) respectively. The notation $Q_0$ refers to the scenario in which no energy flow between the two components of the dark sector are assumed. A lower value of $\Delta A^\prime$ indicates that the corresponding interaction term is favoured by the cosmic chronometers data.  As a term of comparison we exhibit also the result for the $\Lambda$CDM model. The cosmological consequences of these results are explored in Table \ref{tableparams}, where the corresponding values for the present day deceleration parameter, adiabatic speed of sound inside the dark energy fluid, and effective equation of state parameters  for dark energy and dark matter are exhibited. We refer as well to Sect.\ref{secdiscussion} for a discussion about the cosmological meaning of our results. } 
	\label{fig1}
\end{figure}

We exhibit in Table \ref{table33} the best fit values for the model parameters $H_0$, $\Omega_{m0}$, $A$, $B$, $\alpha$,  and $b$ for each dark energy - dark matter interacting model. Table \ref{tableparams} shows the present-day values of the deceleration parameter $ q_{0} $, adiabatic speed of sound  squared for the dark energy fluid $ c_{s}^{2}=\frac{\partial p}{\partial \rho_{de}} = A+\frac{\alpha B}{\rho_{\rm de}^{\alpha+1}} $, and the effective equation of  state parameters for dark energy $\omega=\frac{p}{\rho_{de}}$, and dark matter  $\omega_{eff}=\frac{\Pi}{\rho_{m}}$ for each interacting model. The deceleration parameter is computed using the formula $q=\frac{1+3\omega(1-\Omega_m)}{2}$. We display in Fig.\ref{fig1} the cumulative distribution of the differential area  $\Delta A^{\prime} $, as calculated from (\ref{diffA}), for each cosmological model under investigation.   We remind that the interaction terms between dark energy and dark matter can be found in (\ref{Q1}), ..., (\ref{Q8}) for $Q_1$, ..., $Q_8$ respectively and that the former fluid is pictured according to the equation of state (\ref{chapeq}).

The error bars presented in Table \ref{table33} are found by applying Markov-Chain-Monte-Carlo hammer (emcee)  Bayesian data analysis with flat priors about the mean values we have previously found through the interplay of gaussian reconstruction and  optimization process\footnote{Our python code is an appropriate re-elaboration of the freely available one \href{https://emcee.readthedocs.io/en/stable/tutorials/line/}{\color{blue}{ https://emcee.readthedocs.io/en/stable/tutorials/line/}}} \cite{mcm}. This procedure relies on the assumption that the data in Table \ref{tabledata} come with independent Gaussian errors, and we refer to \cite{jim} for an assessment of this claim.  Then, the error bars in Table \ref{tableparams} are computed from those in Table \ref{table33} via the propagation formula
	\beq
	\label{propagating}
	\sigma_{f(x,y,z,...)}=\sqrt{\left( \frac{\partial f(x,y,z,...)}{\partial x}\right)^2\sigma_x^2 + \left(\frac{\partial f(x,y,z,...)}{\partial y}\right)^2\sigma_y^2 + \left(\frac{\partial f(x,y,z,...)}{\partial z}\right)^2\sigma_z^2 +... } \,.
	\eeq

Moreover,  we have found  that should we replace the gaussian kernel (\ref{gkernel}) with the Mat\'ern one\footnote{It is already known that the Mat\'ern kernel provides a less smooth  reconstructed curve for the function $H(z)$ vs. $z$ \cite[Fig.7]{Melia}  making more problematic for the $\Lambda$CDM model to account for the data at high redshift because the reconstruction delivers a slower increase than the one predicted by the model.  } \cite[Eq.(A.1)]{Melia}
\beq
k\left(z_{1}, z_{2}\right)=\sigma_{f}^{2} {\rm exp} \left( -\frac{3|z_1-z_2|}{l} \right) \left (1+\frac{3|z_1-z_2|}{l}+ \frac{27|z_1-z_2|^2}{7l^2}+\frac{18|z_1-z_2|^3}{7l^3}+\frac{27|z_1-z_2|^4}{35l^4}\right)
\eeq
in the  reconstruction of the cosmic history $H=H(z)$ before performing the optimization process, the estimates of the mean values we have presented in Table \ref{table33} would be affected at most by a 3\% variation. This is less  than both their 1$\sigma$ uncertainty and than the uncertainties which affect the astrophysical data from Table \ref{tabledata} on which the reconstruction is based.

\begin{table}[ht]
	\centering
{	\begin{tabular}{|c|c|c|c|c|c|c|c|}
			\hline 
	& \multicolumn{7}{|c|} {\text {  }}	\\[-0.7em]
			 & \multicolumn{7}{|c|} {\text { Parameters }}   \\
			[2 pt]
			\hline
			&&&&&&&\\[-0.7em]
				$Q_{i}$ & $\Delta A^{\prime}$ & $H_{0}$ & $\Omega_{m 0}$ & $A$ & $B$ & $\alpha$ & $b$ \\
				[2 pt]
			\hline
				&&&&&&&\\[-0.7em]
				$Q_{1}$ & 0.01306 & $66.48_{-10.09}^{+9.97}$ & $0.3107_{-0.0987}^{+0.0994}$ & $-0.3343_{-0.0988}^{+0.1004}$ & $1.8493_{-0.0991}^{+0.1001}$ & $-0.8334_{-0.0099}^{+0.0099}$ & $ 0.0830_{-0.1008}^{+0.0993}$ \\
				[2pt]
			\hline
				&&&&&&&\\[-0.7em]
				$Q_{2}$ & 0.01220 & $66.47_{-9.89}^{+9.88} $& $0.3145_{-0.1004}^{+0.1002} $&$-0.0002_{-0.0999}^{+0.1009} $& $1.6758_{-0.0991}^{+0.1007} $&$-0.9168_{-0.0097}^{+0.0099} $& $0.0837_{-0.1004}^{+0.0993}$ \\
			[2 pt]
			\hline
				&&&&&&&\\[-0.7em]
				$Q_{3}$ & 0.02023 & $66.52_{-10.00}^{+9.90} $ & $ 0.4580_{-0.0988}^{+0.0990} $ & $-0.1951_{-0.1004}^{+0.0993}$ & $2.1950_{-0.0989}^{+0.1000}$ &$ -0 .9100_{-0.0096}^{+0.0098}$ & $0.0827_{-0.1007}^{+0.0980}$ \\
				[2 pt]
			\hline
				&&&&&&&\\[-0.7em]
				$Q_{4}$ & 0.01733 & $ 66.55_{-9.92}^{+10.07}$ & $ 0.2961_{-0.0992}^{+0.1004}$ & $-0.5401_{-0.0999}^{+0.0990}$ & $2.4000_{-0.0985}^{+0.0993}$ & $-0.7500_{-0.0097}^{+0.0099} $ & $0.0854_{-0.0996}^{+0.0986}$\\
				[2 pt]
			\hline
				&&&&&&&\\[-0.7em]
				$Q_{5}$ & 0.01269 & $66.44_{-9.82}^{+10.02}$ & $ 0.3147_{-0.1001}^{+0.0990}$ & $-0.3332_{-0.1000}^{+0.0989}$ & $ 2.0252_{-0.0993}^{+0.0996}$ & $-0.8333_{-0.0100}^{+0.0099}$ & $ 0.0836_{-0.0989}^{+0.0988} $ \\
				[2 pt]
			\hline
				&&&&&&&\\[-0.7em]
				$Q_{6}$ & 0.01275 & $ 66.47_{-9.90}^{+10.04}$ & $0.3159_{-0.0997}^{+0.0992}$ & $-0.3332_{-0.0996}^{+0.0993}$ & $ 2.0246_{-0.0994}^{+0.0996}$ & $-0.8334_{-0.0098}^{+0.0100}$ & $ 0.1670_{-0.0995}^{+0.0999}$ \\
				[2 pt]
			\hline
			&&&&&&&\\[-0.7em]
			$Q_{7}$ & 0.01383 &  $ 66.44_{-9.96}^{+9.92} $& $ 0.3143_{-0.1002}^{+0.0996}$ & $-0.5001_{-0.0985}^{+0.0993}$ & $2.0252_{-0.0997}^{+0.0992} $& $-0.7501_{-0.0099}^{+0.0099}$ & $0.1669_{-0.0994}^{+0.0984}$\\
				[2 pt]
			\hline
			&&&&&&&\\[-0.7em]
			$Q_{8}$ & 0.01168 & $66.51_{-9.83}^{+9.93}$ & $0.3146_{-0.0995}^{+0.0994}$ & $-0.1669_{-0.0982}^{+0.0992}$ & $ 1.3250_{-0.0994}^{+0.0997}$ & $-0.9166_{-0.0100}^{+0.0099}$ & $ 0.1673_{-0.0999}^{+0.0985}$ \\
			[2 pt]
			\hline
			&&&&&&&\\[-0.7em]
			$Q_{0}$ & 0.02157 & $66.54_{-9.92}^{+9.85}$ & $ 0.3188_{-0.0975}^{+0.0988}$ & $-0.5997_{-0.1003}^{+0.0995}$& $1.4409_{-0.0995}^{+0.0984}$ & $-0.0808_{-0.0983}^{+0.0987}$&0
			\\
			[2 pt]
			\hline
		\end{tabular}}
		\caption{The optimal values for the model free parameters $A$, $B$, $\alpha$, $b$,  $H_{0}$ and  $\Omega_{m0}$  for each dark energy - dark matter interacting model. We remind that the equation of state we are adopting for the dark energy is $p=A \rho_{de}-\frac{B}{\rho_{de}^{\alpha}}$, and that $b$ quantifies the strength of interaction between dark energy and dark matter according to the modelings (\ref{Q1}), ..., (\ref{Q8}) for $Q_1$, ..., $Q_8$ respectively. For the case of $Q_0$ appearing in the last row we have set $b=0$, i.e. no interaction between dark energy and dark matter, by assumption. The best models in light of cosmic chronometers data are the ones with a lower value of $\Delta A^\prime$. The  Hubble constant is expressed in units of km/Mpc/s.}
		\label{table33}
	\end{table}

\begin{table}[ht]
	\centering
	{	\begin{tabular}{|c|c|c|c|c|}
			
			\hline
			&&&&\\[-0.7em]$Q_{i}$ & $q_{0}$ & $c_{s}^{2}$ & $\omega$ & $\omega_{e f f}$ \\
			[2 pt]
			\hline
			&&&&\\[-0.7em]$Q_{1}$ &$-0.2640_{-0.1112}^{+0.1127}$& $-0.6715_{-0.1061}^{+0.1077}$&$-0.7389_{-0.1075}^{+0.1090}$&$0.1841_{-0.2236}^{+0.2203}$ \\
		[2 pt]
		\hline
		&&&&\\[-0.7em]
		$Q_{2}$ &$-0.3074_{-0.1338}^{+0.1357}$&$-0.7199_{-0.1298}^{+0.1316}$&$-0.785_{-0.1302}^{+0.1319}$ &$0.0836_{-0.1004}^{+0.0993}$ \\
			[2 pt]
			\hline
			&&&&\\[-0.7em]$Q_{3}$ & $-0.4444_{-0.1124}^{+0.1128}$ & $-1.0747_{-0.1382}^{+0.1387}$ & $-1.1617_{-0.1448}^{+0.1387}$ & $0.1805_{-0.2198}^{+0.2139}$\\
			[2 pt]
			\hline
			&&&&\\[-0.7em]$Q_{4}$ & $-0.3279_{-0.1051}^{+0.1042}$ &$-0.7230_{-0.1016}^{+0.1007}$& $-0.7840_{-0.1023}^{+0.1014}$ & $0.1176_{-0.1160}^{+0.1164}$ \\
			[2 pt]
			\hline
			&&&&\\[-0.7em]$Q_{5}$ & $-0.2981_{-0.1130}^{+0.1119}$ & $-0.7025_{-0.1085}^{+0.1073}$& $-0.7764_{-0.1100}^{+0.1088}$ &$0.1233_{-0.1459}^{+0.1457}$ \\
			[2 pt]
			\hline
			&&&&\\[-0.7em]$Q_{6}$ & $-0.2971_{-0.1122}^{+0.1122}$ & $-0.7029_{-0.1079}^{+0.1079}$ & 	$-0.7768_{-0.1093}^{+0.1094}$ 	& $0.1142_{-0.0680}^{+0.0683}$ \\
			[2 pt]
			\hline
			&&&&\\[-0.7em]$Q_{7}$ & $-0.2279_{-0.1036}^{+0.1044}$ & $-0.6558_{-0.1001}^{+0.1008}$	& $-0.7077_{-0.1007}^{+0.1015}$ 	& $0.2496_{-0.1487}^{+0.1472}$	\\
			[2 pt]
			\hline
			&&&&\\[-0.7em]$Q_{8}$ 	& $-0.3085_{-0.1258}^{+0.1264}$ 	& $-0.7347_{-0.1217}^{+0.1223}$ 	& $-0.7864_{-0.1224}^{+0.1230}$	& $0.0526_{-0.0314}^{+0.0309}$  \\
		[2 pt]
		\hline
		&&&&\\[-0.7em]
		$Q_{0}$  & $-0.1131_{-0.1024}^{+0.1016}$ & $-0.5997_{-0.1003}^{+0.0995}$	& $-0.6000_{-0.1003}^{+0.0995}$ & $0$  \\
		[2 pt]
		\hline
		\end{tabular}}
	\caption{The present-day values of the deceleration parameter $q_{0}$, adiabatic speed of sound squared for the dark energy fluid $c_{s}^{2}=\frac{\partial p}{\partial \rho_{d e}}$, and the effective equation of state parameters for dark energy $\omega=\frac{p}{\rho_{d e}}$ and dark matter $\omega_{e f f}=\frac{\Pi}{\rho_{m}}$ for each cosmological model. These quantities have been computed from those in Table \ref{table33} and the corresponding error bars are found by applying the propagating formula (\ref{propagating}). The expressions for the interaction terms   $Q_1$, ..., $Q_8$ can be found in (\ref{Q1}), ..., (\ref{Q8}) respectively. The model $Q_0$ corresponds to the choice $b=0$, i.e. no interaction between dark energy and dark matter. The best model in light of the cosmic chronometers data is $Q_8$ (see Table \ref{table33}). The uncertainties are hugely affected by the errors on the datapoints at high redshift $z\sim2$ and on the datapoint at $z=0.48$ (see Table \ref{tabledata}). A complete discussion on the cosmological consequences of the results here presented can be found in  Sect.\ref{secdiscussion}.}
	\label{tableparams}
\end{table}


\subsection{Discussion}
\label{secdiscussion}

First of all, it should be noted that the results exhibited in Table \ref{table33} clearly suggest that the simplest one-parameter Chaplygin gas model $p=-\frac{B}{\rho_{de}}$ cannot account for the cosmic chronometers data because we have obtained that the dark energy density should come with a positive power, e.g. $\alpha<0$, in its equation of state. More technically, when insisting to assume such a model, we  could not find any global minimum for the differential area $\Delta A^\prime$ when performing the optimization process. This result should not be naively interpreted as suggesting that a three-parameter model like the Modified  Chaplygin Gas of (\ref{chapeq}) performs better than the Chaplygin Gas just because it involves two more free parameters which can be appropriately tuned, but it is a genuinely physical result. We would like to mention as well that this is not the first time that a negative value of $\alpha$ is estimated: see for example \cite{AA2019,nega} where Planck 2015, type-Ia supernovae, and Hubble parameter data are used. In this paper we found that assuming the dark sector to be composed by two components interacting with each other, our estimates deviate more from the Chaplygin Gas behavior  than the ones in this previous investigation. Furthermore, it should be noted also that  the scenario in which no energy flows between dark energy and dark matter occur is not favoured either, and that the best model is actually the one based on non-linear interactions between dark matter and dark energy as $\sim\rho_m^2/(\rho_m +\rho_{de})$.  More specifically, the optimization process applied to the $\Lambda$CDM model delivers the estimates $H_0=70.17_{-2.75}^{+2.88}$, $\Omega_{m0}=0.2844_{-0.0991}^{+0.1001}$, $q_0= -0.3279_{-0.1051}^{+0.1042}$, with $\Delta A'=0.02249$; it should be noted that already the results reported in Table \ref{table33} implicitly suggest that according to our analysis $\Lambda$CDM is not the favoured model because it can be obtained from ours by fixing $A=-1$, $B=0=b$.  We need also to remark that  the estimates of the cosmological parameters for all the modelings of the interaction terms presented in Tables  \ref{table33}-\ref{tableparams} are degenerate with each other within the 1$\sigma$ interval.

A more transparent physical characterization of the dark energy fluid modeled  according to the Modified Chaplygin Gas (\ref{chapeq}) comes from the study of the following energy conditions \cite{wald}:
\begin{eqnarray}
&&{\rm Null \,\, energy \,\,condition:} \quad \rho+p \geq 0 \,; \\
&&{\rm Weak \,\, energy \,\,condition:} \quad \rho \geq 0\,, \quad \rho+p \geq 0 \,; \\
&&{\rm Dominant \,\, energy \,\,condition:} \quad \rho \geq |p| \,; \\
&&{\rm Strong \,\, energy \,\,condition:} \quad \rho+p \geq 0\,, \quad \rho+3p \geq 0 \,.
\end{eqnarray}
Explicitly they read as:
\begin{eqnarray}
&&{\rm Null \,\, energy \,\,condition:} \quad (1+A)\rho -\frac{B}{\rho^\alpha} \geq 0 \,; \\
&&{\rm Weak \,\, energy \,\,condition:} \quad \rho \geq 0\,, \quad (1+A)\rho -\frac{B}{\rho^\alpha} \geq 0 \,; \\
&&{\rm Dominant \,\, energy \,\,condition:} \quad \rho \geq \Big|A \rho -\frac{B}{\rho^\alpha}\Big| \,;  \\
&&{\rm Strong \,\, energy \,\,condition:} \quad (1+A)\rho -\frac{B}{\rho^\alpha} \geq 0\,, \quad (1+3A)\rho -\frac{3 B}{\rho^\alpha} \geq 0 \,. 
\end{eqnarray}
For example, a phantom energy fluid violates all the null, weak, and strong energy conditions \cite{cald,ss1,ss2,ss3,ss4,ss5}, while a cosmological constant term and a quintessence fluid violate only the strong energy condition \cite{quint1,quint2,quint3}. 
For the case of the Modified Chaplygin Gas analyzed in this paper, considering the present-day value of the effective equation of state parameter $\omega$ from Table \ref{tableparams}, we can conclude that regardless the modeling of the interaction term, but for $Q_3$, only the strong energy condition is violated. This is a remarkable difference between our class of interacting models versus the strengthened dark energy proposal of \cite{Xia} which instead violates also the null and weak energy conditions. 

Since our numerical investigation suggests that $c_s^2<0$, we need to mention that the issue of the stability under small-wavelength perturbations for such configurations was addressed in \cite{corr1,stab2,stab3}, and indeed literature exhibits examples of cosmological applications involving fluids supported by a negative adiabatic speed of sound squared \cite{stab4}.

Combined interpretation of the Planck Power Spectra + Baryon Acoustic Oscillation + Lens in a perturbed flat Friedmann Universe has allowed the reconstruction of the cosmic history of the dark matter equation of state \cite{prl2018} in the framework of so-called \lq\lq Generalized Dark Matter" \cite{prl2018,Wayne}. Our results exhibited in Table  \ref{tableparams} show that cosmic chronometers data favor the  coldest  dark matter equation of state, that is, the one with the  smallest  deviation of the effective equation of state parameter from zero,   although this does not mean the one with the weakest coupling $b$  between the two dark fluids because of the non-linear form of the interaction term. In light of this analysis, the plausible candidates for dark matter should be objects with slower non-relativistic velocities. 
More in general, for all the possible dark energy - dark matter interactions that we have assumed we have obtained a positive value for the parameter $b$ implying that dark energy is decaying into the dark matter in agreement with the Le Ch\^atelier-Braun principle \cite{thermo1}.

\section{conclusion}\label{V}

In this paper, we have shown that also a model selection in light of the cosmic chronometer datasets favors some sort of interaction between dark energy and dark matter beyond the coincidence problem and the Hubble tension issue already extensively investigated in the literature. Another way of interpreting our result would be that in our framework dark matter is not any longer a pressure-less dust fluid but a material with a non-trivial evolving equation of state as in the \lq\lq Generalized Dark Matter" proposal because the interaction term can be recast as an effective contribution to the dark matter pressure.  In fact, the fitting procedure has delivered a nonzero value for the constant quantifying the amount of the energy flow between dark energy and dark matter. Our result seems quite robust because we have explored many possible different realizations of the interaction term beyond the first-order linear approximation by allowing it to depend on several combinations of the dark energy, or on the dark matter amount, or on a combination of them. Similarly,  we have obtained  estimates of $H_0\simeq 66$ km/s/Mpc and $\Omega_m \simeq 0.3$ for all the interaction terms we have assumed. Furthermore, our analysis keeps suggesting that the simpler cosmological constant modeling of the dark energy fluid should be replaced by some sort of evolving field, and actually that also the Generalized Chaplygin Gas scenario (arising in the stringy Born-Infeld theory) should be promoted to the case of the Modified Chaplygin Gas. Moreover, we have also commented that an interaction term between the two cosmic fluids may also avoid the occurrence of a big rip singularity without the need of invoking any mysterious quantum gravity effect because a phantom fluid scenario was ruled out by investigating the energy conditions for the best-fit values of the model parameters. In a set of future works, we will explore more in detail whether our results depend on the particular fluid approach chosen for the modeling of the dark energy.

\subsection*{Acknowledgement}
DG acknowledges support from China Postdoctoral Science Foundation (grant No.2019M661944) and thanks University of Science and Technology of China for hospitality. MA is thanking Prof.Yifu Cai for his helpful suggestions during the project.

{}
\end{document}